\documentclass[conference]{IEEEtran}

\usepackage{amsmath,amsfonts,amssymb,bm}
\usepackage[tight,footnotesize]{subfigure}
\usepackage{graphicx,stfloats}
\usepackage{algorithm}
\usepackage{algorithmic}
\usepackage{cite,url}
\usepackage{cases}
\usepackage{extarrows}
\usepackage{theorem} 
\usepackage{paralist}
\usepackage{psfrag}
\usepackage{auto-pst-pdf}
\usepackage{subfigure}
\usepackage{bm}
\usepackage{float}

\allowdisplaybreaks[4]  
\theorembodyfont{\rmfamily}

\newtheorem{Theorem}{Theorem}

\newcommand{\Lset}{\ensuremath{\mathcal{L}}}
\newcommand{\Cset}{\ensuremath{\mathbb{C}}}
\newcommand{\Eset}{\ensuremath{\mathbb{E}}}
\newcommand{\Rset}{\ensuremath{\mathbb{R}}}

\def\mathLarge#1{\mbox{\Large $#1$}}  

\makeatletter

\newcommand{\Rmnum}[1]{\expandafter\@slowromancap\romannumeral #1@}
\makeatother

\IEEEoverridecommandlockouts

\begin{document}

\title{A Multi-cell MMSE Precoder for Massive MIMO Systems and New Large System Analysis}
\author{\IEEEauthorblockN{Xueru~Li$^{\dagger}$, Emil~Bj{\"o}rnson$^{*}$, Erik G. Larsson$^{*}$, Shidong~Zhou$^{\dagger}$ and Jing~Wang$^{\dagger}$}
\IEEEauthorblockA{$^{\dagger}$ State Key Laboratory on Microwave and Digital Communications\\
Tsinghua National Laboratory Information Science and Technology\\
Department of Electronic Engineering, Tsinghua University, Beijing 100084, China.\\
$^{*}$ Department of Electrical Engineering (ISY), Link{\"o}ping University, SE-58183 Link{\"o}ping, Sweden.\\
Email: xueruli1206@163.com, emil.bjornson@liu.se.}
\thanks{The work is partially supported by National Basic Research Program (2012CB316000), National S\&T Major Project (2014ZX03003003-002), National Natural Science Foundation of China (61201192), National High Technology Research and Development Program of China (2014AA01A703), Tsinghua-HUAWEI Joint Research \& Development on Soft Defined Protocol Stack, ICRI-MNC, Tsinghua-Qualcomm Joint Research Program, Keysight Technologies, Inc., ELLIIT, the CENIIT project 15.01 and FP7-MAMMOET.}\vspace{-2ex}}

\maketitle
\begin{abstract}
In this paper, a new multi-cell MMSE precoder is proposed for massive MIMO systems. We consider a multi-cell network where each cell has $K$ users and $B$ orthogonal pilot sequences are available, with $B = \beta K$ and $\beta \ge 1$ being the pilot reuse factor over the network. In comparison with conventional single-cell precoding which only uses the $K$ intra-cell channel estimates, the proposed multi-cell MMSE precoder utilizes all $B$ channel directions that can be estimated locally at a base station, so that the transmission is designed spatially to suppress both parts of the inter-cell and intra-cell interference. To evaluate the performance, a large-scale approximation of the downlink SINR for the proposed multi-cell MMSE precoder is derived and the approximation is tight in the large-system limit. Power control for the pilot and payload, imperfect channel estimation and arbitrary pilot allocation are accounted for in our precoder. Numerical results show that the proposed multi-cell MMSE precoder achieves a significant sum spectral efficiency gain over the classical single-cell MMSE precoder and the gain increases as $K$ or $\beta$ grows. Compared with the recent M-ZF precoder, whose performance degrades drastically for a large $K$, our M-MMSE can always guarantee a high and stable performance. Moreover, the large-scale approximation is easy to compute and shown to be accurate even for small system dimensions.
\end{abstract}

\section{Introduction} \label{intro}
Massive multiple-input-multiple-ouput (MIMO) is an exciting new multi-user MIMO technology that has gained traction in recently years~\cite{Marzetta10,Rusek13}. The idea is to employ an array comprising hundreds of antennas at each base station (BS) and serve tens of users simultaneously. The system spectral efficiency (SE) can be improved by an order of magnitude without consuming extra spectral resource~\cite{Marzetta10, Rusek13, Larsson14,bjornson2014massive}. By using simple coherent linear precoders such as zero forcing, the intra-cell interference and receiver noise can be averaged out in the limit of an infinite number of BS antennas, and the only remaining performance bottlenecks are pilot contamination and distortion noise from hardware impairments~\cite{Marzetta10,bjornson2013massive}. The uplink and downlink transmit powers can also be reduced by an order of magnitude since a comparable array gain can be harnessed by the phase-coherent processing~\cite{Hien13}. Furthermore, in time division duplex (TDD) mode, the channel estimation overhead scales linearly with the number of users, instead of the number of BS antennas, which allows for adding extra antennas without affecting the estimation overhead~\cite{Marzetta2006}. These features make massive MIMO one of the key technologies for the next generation wireless communication systems.

In the downlink, the most common linear precoding schemes are matched filtering (MF), zero forcing (ZF) and minimum mean square error (MMSE).\footnotemark{ }We consider a multi-cell network with $B$ orthogonal uplink pilot sequences and $K$ users in each cell, where $B = \beta K$ and $\beta\ge 1$ is called the pilot reuse factor (i.e., only $1/\beta$ of the cells use the same set of pilots). In conventional massive MIMO systems, each BS estimates the $K$ intra-cell channels by listening to the pilot signalling from its own cell, and then constructs user-specific precoders based on the channel estimates to limit intra-cell interference~\cite{Marzetta10,Hoydis2013,Yang2013performance}. When $\beta > 1$, however, the BS can locally estimate $B$, instead of $K$, channel directions by also listening to the pilot signalling from all cells. Thereby, the BS is able to select its precoders to actively suppress also the inter-cell interference, since its $K$ serving users only occupy $K$ out of the $B$ channel directions. A downlink multi-cell precoding scheme is proposed in~\cite{Jose2011}, which brings a notable gain over single-cell processing. However, this scheme does not account for optimized pilot allocation which, as shown in~\cite{bjornson2014massive}, is an important way to suppress pilot contamination and achieve high system SE in massive MIMO deployments. Moreover, no closed-form performance analysis is provided in~\cite{Jose2011}.
\footnotetext{A special case of the MMSE precoder is the regularized ZF precoder, which is obtained when all the users in a cell have equal pathlosses.}

In this paper, we propose a new multi-cell MMSE precoder and derive a large-scale approximation of the downlink SINR, which is tight the large-system limit. Power control for the pilot and payload, imperfect channel estimation and arbitrary pilot allocation are all accounted for in our scheme. By utilizing all the channel directions that can be locally estimated at a BS, the proposed multi-cell MMSE precoder can not only suppress intra-cell interference, but also actively reduce the interference caused to other cells. Numerical results are provided to show that a significant sum SE gain can be obtained by the proposed precoder over the single-cell MMSE precoder and the gain becomes more significant as $K$ or $\beta$ increase. Compared to the multi-cell ZF precoder from~\cite{bjornson2014massive}, whose performance degrades drastically for larger $K$ or $\beta$, our M-MMSE precoder provides a high and stable performance even for very large $K$. Furthermore, the large-scale approximation of the SINR is easy to compute and shown to be a very accurate approximation even for small system dimensions.

\textit{Notations}: The trace, conjugate, transpose, Hermitian transpose and matrix inverse operators are denoted by $\rm{tr}\left(\cdot \right)$,  $\left(\cdot \right)^{*}$, $\left(\cdot \right)^T$, $\left(\cdot \right)^H$ and $\left(\cdot \right)^{-1}$, respectively.

\section{System Model and Precoder Design} \label{system model}
We consider a massive MIMO cellular network operating according to a synchronous TDD protocol. Each cell is assigned an index in the cell set $\Lset$, and the cardinality $\left| \Lset \right|$ is the number of cells. The BS in each cell is equipped with $M$ antenna elements and serves $K$ single-antenna users within each coherence block. Assume that this time-frequency block consists of $T_c$ seconds and $W_c$ Hz, such that $T_c$ is smaller than the coherence time of all users and $W_c$ is smaller than the coherence bandwidth of all users. This leaves room for $S=T_c \times W_c$ transmission symbols per block, and the channels of all users remain constant within each block. Let ${{\bf{h}}_{jlk}}$ denote the channel response from user $k$ in cell $l$ to BS $j$ within a block and assume that it is a realization from a zero-mean circularly symmetric complex Gaussian distribution:

\begin{equation} \label{channel model}
{{\bf{h}}_{jlk}} \sim {\cal{CN}}\left( {{\bf 0},{d_j}\left( {{{\bf{z}}_{lk}}} \right){{\bf{I}}_M}} \right).
\end{equation}
The vector ${\bf{z}}_{lk} \in \Rset ^2$ is the geographical position of user $k$ in cell $l$ and ${{d_j}\left( {{{\bf{z}}}} \right)}$ accounts for the channel attenuation (e.g., path loss and shadowing) related to some user position $\bf{z}$. Since the user positions change relatively slowly, ${{d_j}\left( {{{\bf{z}}_{lk}}} \right)}$ is assumed to be known at BS $j$ for all $l$ and all $k$. 

In the TDD protocol, the channels are estimated from uplink pilot signaling and utilized for downlink precoding by exploiting channel reciprocity. 
Thus, the channel estimation is first discussed to lay a foundation for the precoder design.

\subsection{Uplink Channel Estimation} \label{channel estimation}
In the channel estimation phase, the collective received signal at BS $j$ is denoted as ${{\bf{Y}}_j} \in {\Cset^{M \times B}}$ where $B$ is the length of a pilot sequence (and thus the number of orthogonal pilot sequences available in the network). Then ${\bf Y}_j$ can be expressed as

\begin{equation} \label{estimationsignal}
{{\bf{Y}}_j} = \sum\limits_{l \in {\cal L}} {\sum\limits_{k = 1}^K {\sqrt {{p_{lk}}} {{\bf{h}}_{jlk}}{\bf{v}}_{{i_{lk}}}^T} }  + {{\bf{N}}_j},
\end{equation}
where ${\bf{h}}_{jlk}$ is the channel defined in~(\ref{channel model}) and $p_{lk}$ is the power control coefficient for the pilot of user $k$ in cell $l$. The matrix ${{\bf{N}}_j}$ contains independent elements which follow a complex Gaussian distribution with zero mean and variance $\sigma ^2$. We assume that all pilot sequences originate from a predefined orthogonal pilot book, defined as ${\cal{V}}= \left\{ {{{\bf{v}}_1},\ldots,{{\bf{v}}_B}} \right\}$, where
\begin{equation}
{\bf{v}}_{{b_1}}^H{{\bf{v}}_{{b_2}}} = \left\{ \begin{array}{ll}
B,  & {b_1} = {b_2},\\
0,  & {b_1} \ne {b_2},
\end{array} \right.
\end{equation}
and let ${i_{lk}} \in \left\{ 1,\ldots,B\right\}$ denote the index of the pilot sequence used by user $k$ in cell $l$. Arbitrary pilot allocation is supported and the relation between $B$ and $K$ is denoted by $B = \beta K$, where $\beta \ge 1$ is called the pilot reuse factor. If the pilots are allocated wisely between the cells, a larger $\beta$ brings higher estimation quality and a lower level of pilot contamination.

Based on the received pilot signal in~(\ref{estimationsignal}), the MMSE estimate of the uplink channel ${{\bf{h}}}_{jlk}$ is
\begin{equation} \label{estimation}
{\hat{\bf h}}_{jlk} = \sqrt{p_{lk}} d_j\left( {\bf z}_{lk}\right){\bf Y}_j \left({\bf \Psi}_j^{*}\right)^{-1} {\bf v}_{i_{lk}}^{*},
\end{equation}
where ${{\bf{\Psi }}_j}$ is the covariance matrix of the vectorized received signal ${\rm{vec}}\left({\bf Y}_j\right)$ and is given by
\begin{equation}
{{\bf{\Psi }}_j} = \sum\limits_{\ell \in {\cal L}} {\sum\limits_{m = 1}^K {p_{{\ell m}}} d_j\left( {\bf z}_{\ell m}\right){\bf{v}}_{{i_{\ell m}}}{\bf{v}}_{{i_{\ell m}}}^H} + \sigma ^2{{\bf{I}}_B}.
\end{equation}
Then according to the orthogonality principle of MMSE estimation, the covariance matrix of the estimation error ${\tilde {\bf h}}_{jlk} ={{\bf{h}}_{jlk} - {\hat{\bf h}}_{jlk}}$ is given by
\begin{eqnarray}\label{errormatrix}
{{\bf{C}}_{jlk}} &=& {\Eset}\left\{  {{\tilde {\bf{h}}}_{jlk} { {{\tilde {\bf{h}}}_{jlk}^H}}} \right\} \nonumber \\
&=&d_j \left( {{{\bf{z}}_{lk}}} \right)  \left( {1 - p_{lk} d_j \left( {{{\bf{z}}_{lk}}} \right){\bf{v}}_{{i_{lk}}}^H{\bf{\Psi }}_j^{ - 1}{{\bf{v}}_{{i_{lk}}}}} \right){{\bf{I}}_M} \nonumber \\
&=&d_j \left( {{{\bf{z}}_{lk}}} \right)  \left( 1 - p_{lk} d_j \left( {{{\bf{z}}_{lk}}}  \right) {\alpha _{ji_{lk}}}B\right){{\bf{I}}_M},
\end{eqnarray}
where $\alpha _{ji_{lk}}= \frac{1}{B} {\bf v}_{i_{lk}}^H {\bf \Psi}_j^{-1} {\bf v}_{i_{lk}}$ and can be simplified as
\begin{equation} \label{alpha}
\alpha _{ji_{lk}}=\frac{1}{{\sum\nolimits_{\ell  \in {\cal L}} {\sum\nolimits_{m = 1}^K { p_{\ell m}{{{d_j}\left( {{{\bf{z}}_{\ell m}}} \right)}}{\bf{v}}_{{i_{lk}}}^H{{\bf{v}}_{{i_{\ell m}}}}} }  + \sigma ^2}}.
\end{equation}

As pointed out in~\cite{bjornson2014massive}, the term ${\bf Y}_j \left({\bf \Psi}_j^{*}\right)^{-1}{\bf v}_{i_{lk}}^{*}$ in~(\ref{estimation}) depends only on which pilot sequence that user $k$ in cell $l$ uses. Consequently, users who use the same pilot sequence have parallel estimated channels at each BS, while only the amplitudes are different. To show this explicitly, define the $M \times B$ matrix
\begin{eqnarray} \label{H_v}
{{{\hat {\bf H}}}_{\mathcal{V},j}}=\left[ {\hat{\bf h}}_{{\cal V},j1},...,{\hat{\bf h}}_{{\cal V},jB} \right] = {\bf Y}_j \left({\bf \Psi}_j^{*}\right)^{-1} \left[{\bf v}_1^{*},...,{\bf v}_B^{*}\right],
\end{eqnarray}
so that the MMSE estimate in~(\ref{estimation}) can be expressed as
\begin{equation} \label{estimation2}
{\hat {\bf h}}_{jlk} = \sqrt{p_{lk}} d_j \left( {{{\bf{z}}_{lk}}} \right){{{\hat {\bf H}}}_{{\cal V},j}}{{\bf{e}}_{{i_{lk}}}},
\end{equation}
where ${{\bf{e}}_i}$ denotes the $i$th column of the identity matrix ${\bf I}_B$. The fact that users with the same pilot have parallel estimated channels is used herein to derive a new SE expression.

\newcounter{TempEqCnt}
\setcounter{TempEqCnt}{\value{equation}}
\setcounter{equation}{14}

\begin{figure*}[ht]
 \vspace{-2ex}
\begin{equation}\label{sinr_dl}
\eta_{jk} = \frac{{\varrho_{jk}} \left| \Eset_{\left\{\bf h\right\}}\left\{{\bf h}_{jjk}^H {\bf w}_{jk} \right\}\right|^2 }{\sum\limits_{l \in {\cal L}}  \sum\limits_{m=1}^K {\varrho_{lm}} \Eset_{\left\{\bf h\right\}}\left\{ \left| {\bf h}_{ljk}^H {\bf w}_{lm}\right|^2 \right\} - {\varrho_{jk}} \left| \Eset_{\left\{\bf h\right\}}\left\{{\bf h}_{jjk}^H {\bf w}_{jk} \right\}\right|^2  + \sigma^2 }.
\end{equation}
\hrule
 \vspace{-2ex}
\end{figure*}
\setcounter{equation}{\value{TempEqCnt}}

Notice that the estimated channel ${\hat{\bf h}}_{jlk}$ is also a zero-mean Gaussian vector, and its covariance matrix ${{\bf{\Phi }}_{jlk}}$ is
\begin{equation} \label{covariance}
{{\mathbf{\Phi }}_{jlk}} = d_j \left( {{{\bf{z}}_{lk}}} \right){\bf I}_M - {\bf C}_{jlk} = {p_{lk}} d_j^2\left( {{{\bf{z}}_{lk}}}\right){\alpha _{ji_{lk}}}B {\bf I}_M.
\end{equation}
Define the covariance matrix of ${\hat {\bf h}}_{{\cal V},ji}$ as ${ \tilde {\bf \Phi}}_{{\cal{V}},ji}$. Then according to~(\ref{estimation2}) and~(\ref{covariance}), ${\tilde {\bf \Phi}}_{{\cal V},j{i}} = \alpha_{ji}B {\bf I}_M$.

\subsection{Downlink Multi-cell MMSE Precoder}
During the downlink payload transmission, the collective received signal at user $k$ in cell $j$ can be expressed as
\begin{equation}
y_{jk} = \sum\limits_{l \in {\cal L}} {\bf h}_{ljk}^H \sum\limits_{m=1}^K \sqrt{\varrho}_{lm}{\bf w}_{lm} s_{lm} + n_{jk},
\end{equation}
where ${\bf w}_{lm}\in \mathbb{C}^{M \times 1}$ is the precoder associated with user $m$ in cell $l$, $s_{lm} \sim {\cal{CN}} \left(0,1 \right)$ is the corresponding payload data symbol, $\varrho_{lm}$ is the downlink transmit power coefficient, and  $n_{jk}\sim {\cal{CN}} \left(0,1 \right)$ is additive white Gaussian noise (AWGN).

In~\cite{xueru2015UL}, we propose a multi-cell MMSE (M-MMSE) detector for the uplink of massive MIMO. In contrast to conventional detectors that only utilizes the $K$ intra-cell channels, our proposed M-MSME detector utilizes all $B$ estimated channel directions in the matrix ${\hat {\bf H}}_{{\cal V},j}$ from~(\ref{H_v}). Since the $K$ users only occupy $K$ out of the $B$ channel directions, our M-MMSE detector can suppress both intra and inter-cell interference, whereas the conventional detectors can only suppress intra-cell interference. The proposed M-MMSE detector maximizes the uplink SINR under general conditions, and notable SE gains are achieved over conventional single-cell MMSE and the multi-cell ZF. The M-MMSE detector from~\cite{xueru2015UL} is given by
\begin{equation} \label{detector2}
{\bf{g}}_{jk}^{\rm{M-MMSE}}= {\left( {{{{\hat {\bf H}}}_{\mathcal{V},j}}{{\bf{\Lambda }}_j}{\hat {\bf H}}_{\mathcal{V},j}^H + \left( {{\sigma ^2} + {\varphi_j}} \right){{\bf{I}}_M}} \right)^{ - 1}}{\hat {\bf h}}_{jjk},
\end{equation}
where ${{\bf{\Lambda }}_j} = \sum\limits_{l \in L} {\sum\limits_{k = 1}^K \tau_{lk} p_{lk}d_j^2\left({\bf z}_{lk}\right) } {{\bf{e}}_{{i_{lk}}}}{\bf{e}}_{{i_{lk}}}^H$ and $\tau_{lk}$ is the uplink transmit payload power from user $k$ in cell $l$. Moreover, ${\varphi_j} = \sum\limits_{l \in {\cal L}} {\sum\limits_{k = 1}^K {\tau_{lk}d_j\left({\bf z}_{lk}\right) \left(1-p_{lk}d_j\left({\bf z}_{lk}\right) \alpha_{ji_{lk}}B \right) }}$, where $\alpha_{ji_{lk}}$ is defined in~(\ref{alpha}).

Recently,~\cite{bjornson2014massive} established an uplink-downlink duality for massive MIMO systems which proves that the uplink SEs can also be achieved in the downlink if each downlink precoder is a scaled version of the corresponding uplink detector. As the M-MMSE detector is the state-of-the-art uplink method, we propose to apply the same methodology for downlink precoding. The downlink M-MMSE precoder is constructed as
\begin{equation} \label{m_precoder}
{\bf w}_{jk}^{\rm{M-MMSE}} = \frac{ {\bf g}_{jk}^{\rm{M-MMSE}}}{\sqrt{\lambda_{jk}}},
\end{equation}
where $\lambda_{jk} = \Eset\{\|{\bf g}_{jk}^{\rm{M-MMSE}} \|^2 \}$ normalizes the average transmit power for the user $k$ in cell $j$ to $\Eset\{\| \sqrt{\varrho_{lm}}{\bf w}_{jk}s_{lm}\|^2 \}=\varrho_{lm}$. Since there are no downlink pilots in our TDD protocol, the users do not know the current channels but can only learn the average channel gain, $\sqrt{\varrho_{jk}} \Eset_{{\bf h}}\{{\bf h}_{jjk}^H {\bf w}_{jk}\}$, and the total interference variance. Therefore, the ergodic achievable downlink SE
\begin{equation}\label{rate_dl}
R_{jk} =  \left(1-\frac{B}{S} \right) \log_2\left( 1+ \eta_{jk}\right)
\end{equation}
can be obtained at user $k$ in cell $l$ as in~\cite{Hoydis2013,bjornson2014massive}, where $(1-\frac{B}{S})$ is the pilot overhead and $\eta_{jk}$ is given at top of this page.

The downlink SINR in~(\ref{sinr_dl}) holds for any linear precoding scheme. This SE is achieved by treating $\Eset_{\{\bf h\}}\{{\bf h}_{jjk}^H {\bf w}_{jk}\}$ as the true channel in the receiver, and treating interference and channel variations as worst-case Gaussian noise. Thus, $R_{jk}$ is a lower bound on the downlink ergodic capacity.

By utilizing all the available estimated directions, the M-MMSE precoder suppresses both intra-cell interference and the interference caused to other cells, and thus a higher SINR can be expected by our precoder than conventional single-cell precoders, at least for an appropriate power control (as prescribed by the uplink-downlink duality concept in~\cite{bjornson2014massive}). In the next section, a large-scale approximation of the downlink SINR in~(\ref{sinr_dl}) is derived. Note that in~\cite{Jose2011}, the authors also propose a multi-cell MMSE precoder, which brings a notable gain over single-cell processing. But it does not accounted for arbitrary or optimized pilot allocation which, as shown in~\cite{bjornson2014massive}, is an important way to suppress pilot contamination and achieve high system SE in massive MIMO. Moreover, no closed-form performance analysis is provided in~\cite{Jose2011}.

\section{Asymptotic Analysis} \label{asymptotic analysis}
In this section, performance analysis is conducted for the proposed M-MMSE precoder. The large-system limit is considered, where $M$ and $K$ go to infinity while keeping the ratio ${K \mathord{\left/  {\vphantom {K M}} \right. \kern-\nulldelimiterspace} M} $ finite. In what follows, the notation $M \to \infty$ refers to $K$, $M \to \infty$ such that $\lim {\sup _M}{K \mathord{\left/  {\vphantom {K M}} \right. \kern-\nulldelimiterspace} M} < \infty $ and $\lim {\inf _M}{K \mathord{\left/  {\vphantom {K M}} \right. \kern-\nulldelimiterspace} M} >0$.\footnotemark{ }Since $B$ scales with $K$ for a fixed $\beta$, $\lim {\sup _M}{B \mathord{\left/  {\vphantom {K M}} \right. \kern-\nulldelimiterspace} M} < \infty $ and $\lim {\inf _M}{B \mathord{\left/  {\vphantom {K M}} \right. \kern-\nulldelimiterspace} M} >0$ also hold for $B$. The results should be understood in the way that, for each set of system dimension parameters $M$, $K$ and $B$, we provide an approximative expression for the SINR, and the expression is tight as $M$, $K$ and $B$ grow large. In what follows, the notation $\xrightarrow[M \to \infty]{}$ denotes the convergence of a deterministic sequence.
\footnotetext{The limit superior of a sequence $x_n$ is defined by $\lim {\sup _n}{x_n}\triangleq \mathop {\lim }\limits_{n \to \infty } \left( {\sup \left\{ {{x_m}:m \geqslant n} \right\}} \right)$; the limit inferior is defined as $\lim{\inf_n}{x_n}\triangleq \mathop {\lim }\limits_{n \to \infty } \left( {\inf \left\{ {{x_m}:m \geqslant n} \right\}} \right)$.}

Before we continue with our performance analysis, two useful results from large random matrix theory are first recalled. All vectors and matrices should be understood as sequences of vectors and matrices of growing dimensions.

\setcounter{TempEqCnt}{\value{equation}}
\setcounter{equation}{24}
\begin{figure*}[ht]
 \vspace{-2ex}
\begin{equation} \label{sinr_dl_determ}
{{\bar \eta }_{jk}} = \frac{\varrho_{jk} p_{jk} d_j^2\left({\bf z}_{jk} \right)\frac{\delta _{jk}^2}{\vartheta_{jk}^{''}}}{{p_{jk}\sum\limits_{\left( {l,m} \right) \ne \left( {j,k} \right),{i_{_{lm}}} = {i_{jk}}} {\varrho_{lm}d_l^2\left({\bf z}_{jk} \right)\frac{\delta_{lm}^2}{\vartheta_{lm}^{''}}}  + \sum\limits_{{i_{_{lm}}} \ne {i_{jk}}} { \varrho_{lm} d_l\left({\bf z}_{jk} \right)\frac{\mu_{ljkm}}{M \vartheta_{lm}^{''}}}  + \frac{{{\sigma ^2}}}{M}}}.
\end{equation}
\hrule
 \vspace{-2ex}
\end{figure*}
\setcounter{equation}{\value{TempEqCnt}}

\subsection{Useful theorems} \label{theorems}
\begin{Theorem} (Theorem 1 in \cite{Wagner2012}): \label{theorem1}
Let ${\bf D} \in \Cset ^{M \times M}$ be deterministic and ${\bf H} \in \Cset ^{M \times B}$ be random with independent column vectors ${\bf h}_b \sim {\cal {CN}} \left(0, \frac{1}{M}{\bf R}_b \right)$. Assume that $\bf D$ and the matrices ${\bf R}_b \left( b=1,...,B\right)$, have uniformly bounded spectral norms (with respect to $M$). Then, for any $\rho > 0$,
\setcounter{equation}{15}
\begin{equation}
\frac{1}{M} {\rm{tr}}\left({\bf D} \left( {\bf {HH}}^H +\rho {\bf I}_M \right)^{-1}\right) - \frac{1}{M} {\rm{tr}}\left({\bf D}{\bf T}\left( \rho \right)\right) \xrightarrow[M \to \infty]{a.s.} 0,
\end{equation}
where ${\bf T}\left( \rho \right) \in \Cset^{M \times M}$ is defined as
\begin{equation}
{\bf T}\left( \rho \right) = \left( \frac{1}{M}\sum\limits_{b=1}^B \frac{{\bf R}_b}{1+\delta_b \left( \rho\right)} +\rho{\bf I}_M\right)^{-1}
\end{equation}
and the elements of ${\bm {\delta }}\left( \rho  \right) \buildrel \Delta \over = {\left[ {{\delta _1}\left( \rho  \right),...,{\delta _B}\left( \rho  \right)} \right]^T}$ are defined as $\delta_b \left(\rho \right)=\lim_{t \to \infty}\delta_b^{\left(t \right)}\left( \rho \right), b=1,...,B$, where
\begin{equation}
\delta_b^{\left(t \right)}\left( \rho \right) = \frac{1}{M} {\rm{tr}} \left( {\bf R}_b \left( \frac{1}{M} \sum\limits_{j=1}^B \frac{{\bf R}_j}{1+ \delta_j^{\left(t-1\right)} \left(\rho \right)} +\rho{\bf I}_N\right)^{-1}\right)
\end{equation}
for $t=1,2,\ldots,$ with initial values $\delta_b^{\left(0 \right)} = 1/\rho$ for all $b$.
\end{Theorem}

\begin{Theorem}(see \cite{Wagner2012}) \label{theorem2}
Let ${\bf {\Theta}} \in \Cset^{M \times M}$ be Hermitian nonnegative definite with uniformly bounded spectral norm (with respect to $M$). Under the same conditions for ${\bf D}$ and $\bf H$ as in Theorem~\ref{theorem1},
\begin{gather}
\frac{1}{M} {\rm{tr}}\left({\bf D} \left( {\bf {HH}}^H + \rho {\bf I}_M \right)^{-1} {\bf \Theta}\left( {\bf {HH}}^H + \rho {\bf I}_M \right)^{-1} \right) \nonumber \\
- \frac{1}{M} {\rm{tr}}\left({\bf D}{\bf T}'\left( \rho \right) \right) \xrightarrow[M \to \infty]{a.s} 0
\end{gather}
where ${\bf T}'\left( \rho \right) \in \Cset^{M \times M}$ is defined as
\begin{equation}
{\bf T}'\left( \rho \right) ={\bf T}\left( \rho \right) {\bf \Theta} {\bf T}\left( \rho \right) +{\bf T}\left( \rho \right) \frac{1}{M} \sum\limits_{b=1}^B \frac{{\bf R}_b \delta'_b\left(\rho \right)}{\left(1+ \delta_b\left(\rho \right)\right)^2}{\bf T}\left( \rho \right).
\end{equation}
${\bf T}\left( \rho \right)$ and ${\bm \delta}\left( \rho  \right)$ are given by Theorem~\ref{theorem1}, and ${\bm{\delta }}'\left( \rho  \right)= \left[{\delta}'_1\left( \rho  \right),...,{\delta}'_B\left( \rho  \right) \right]^T$ is calculated as
\begin{equation}
{\bm \delta}' \left( \rho  \right)= \left({\bf I}_B - {\bf J}\left( \rho \right) \right)^{-1} {\bf v}\left( \rho \right)
\end{equation}
where ${\bf J}\left( \rho \right)$ and $ {\bf v}\left( \rho \right)$ are defined as
\begin{equation}
\left[ {\bf J}\left( \rho \right)\right]_{bl} = \mathLarge {\frac{ \frac{1}{M} {\rm{tr}} \left({\bf R}_b {\bf T}\left( \rho \right) {\bf R}_l {\bf T}\left( \rho \right)\right)} {M \left(1+\delta_l\left( \rho \right) \right)^2 }},  1 \le b,l \le B \\
\end{equation}
\begin{equation}
\left[ {\bf v}\left( \rho \right)\right]_{b} = \frac{1}{M}{\rm{tr}}\left({\bf R}_b{\bf T}\left( \rho \right) {\bf \Theta}{\bf T}\left( \rho \right)\right),  1 \le b \le B.
\end{equation}
\end{Theorem}

\subsection{A Large-Scale Approximation of the SINR} \label{deterministic sinr} 
In what follows, we derive the large-scale approximation ${\bar \eta}_{jk}$ of ${\eta}_{jk}$ with the M-MMSE precoder such that
\begin{equation}
{\bar \eta}_{jk}-{\eta}_{jk} \xrightarrow[M \to \infty]{} 0.
\end{equation}
\begin{Theorem} \label{theorem4}
For the downlink MMSE precoder in~(\ref{m_precoder}), we have $\eta_{jk} - {{\bar \eta }_{jk}} \xrightarrow[M \to \infty]{}0$, where $\bar{\eta}_{jk}$ is given at top of this page
with
\setcounter{equation}{25}
\begin{flalign}
& \delta_{jk}=\frac{1}{M} {\rm{tr}} \left({\tilde {\bf \Phi}}_{{\cal V},ji_{jk}} {{\bf T}}_j\right), &
\end{flalign}
\begin{flalign}
\mu_{ljkm}&=\frac{1}{M}{\rm{tr}} \left({{{{\mathbf{ T}}}_{lm}^{'}}}\right) \nonumber \\
 \begin{split}
&- p_{jk} d_l\left({\bf z}_{jk}\right)\gamma_{li_{jk}}\vartheta_{ljkm}^{'}\vartheta_{ljk} \frac{2+\gamma_{li_{jk}} \vartheta_{ljk} }{ \left(1+ \gamma_{li_{jk}} \vartheta_{ljk}\right)^2 }, \quad
 \end{split}
\end{flalign} \vspace{-2ex}
\begin{flalign}
&\vartheta_{ljk}=\frac{1}{M}{\rm{tr}}\left( {\tilde {\bf \Phi}}_{{\cal V},li_{jk}}{\bf T}_l\right),&
\end{flalign}
\begin{flalign}
&\vartheta_{ljkm}^{'}=\frac{1}{M}{\rm{tr}}\left( {\tilde {\bf \Phi}}_{{\cal V},li_{jk}}{\bf T}_{lm}^{'}\right),&
\end{flalign}
\begin{flalign}
&\vartheta_{lm}^{''}=\frac{1}{M}{\rm{tr}}\left( {\tilde {\bf \Phi}}_{{\cal V},li_{lm}}{\bf T}_{lm}^{''}\right), &
\end{flalign}
where
\begin{enumerate}
\item ${\bf T}_l = {\bf T}_l\left(\alpha \right)$ and ${\bm{\delta }}\left( \alpha \right) \buildrel \Delta \over = {\left[ {{\delta _1},...,{\delta _B}} \right]^T}$ are given by Theorem~\ref{theorem1} for $\alpha = \frac{\sigma^2 + \varphi_l}{M}$ and ${\bf R}_b =\gamma_{lb}{ \tilde {\bf \Phi}}_{{\cal{V}},lb}$.  
\item ${\bf T}_{lm}^{'} = {\bf T}_{lm}^{'}\left(\alpha \right)$ and ${\bm{\delta }}'\left( \alpha  \right)= \left[{\delta}'_1,...,{\delta}'_B \right]^T$ are given by Theorem~\ref{theorem2} for $\alpha = \frac{\sigma^2 +\varphi_l}{M}$, ${\bf \Theta}={\tilde{\bf \Phi}}_{{\cal V},li_{lm}}$, and ${\bf R}_b =\gamma_{lb}{\tilde {\bf \Phi}}_{{\cal{V}},lb}$. 
\item ${\bf T}_{lm}^{''}={\bf T}_{lm}^{''} \left(\alpha \right)$ and ${\bm{\delta }}'\left( \alpha  \right)= \left[{\delta}'_1,...,{\delta}'_B \right]^T$ are given by Theorem~\ref{theorem2} for $\alpha = \frac{\sigma^2 + \varphi_l}{M}$, ${\bf \Theta}={\bf I}_M $, and ${\bf R}_b =\gamma_{lb}{\tilde {\bf \Phi}}_{{\cal{V}},lb}$.  
\end{enumerate}
\end{Theorem}
\noindent\emph{Proof:} See the Appendix B and C of~\cite{xueru2015}. \hfill{$\blacksquare$}

By utilizing Theorem~\ref{theorem4}, the ergodic SE $R_{jk}$ in~(\ref{rate_dl}), after dropping the prelog factor, converges to ${\bar R}_{jk}=\log_2 (1+{\bar \eta_{jk}})$ in the large-system limit. Therefore, a large-scale approximation of the downlink ergodic SE is provided by $(1-\frac{B}{S}){\bar R}_{jk}$. This approximation is easy to compute and, as shown in Section~\ref{simulation}, is very accurate even at small system dimensions.

\section{Simulation Results} \label{simulation}
In this section, we illustrate the accuracy and usefulness of the analytical contributions for a symmetric hexagonal network topology. We consider the classic 19-cell-wrap-around structure to avoid edge effects and guarantee the same SE in all cells~\cite{bjornson2014massive,xueru2015}. Each hexagonal cell has a radius of $r = 500$ meters. To achieve a symmetric pilot allocation network, the non-universal reuse factor can be $\beta \in \left\{1,3,4,7,\cdots\right\}$, and the pilots are then allocated randomly within each cell.

User locations are generated independently and randomly in the cells by following uniform distributions, but the distance to the serving BS is at least $0.14r$. For each user location ${\bf z} \in \Rset^{2}$, a classic pathloss model is considered, where the variance of channel attenuation is $d_j\left( {\bf z} \right) = \frac{C}{\left\| {\bf z} - {\bf b}_j\right\|^{\kappa}}$. Here ${\bf b}_j \in \Rset^{2}$ is the location of BS in cell $j$, $\kappa$ is the pathloss exponent, and $\left\|\cdot \right\|$ denotes the Euclidean norm. $C>0$ is independent shadow fading with $10\log_{10}\left(C \right)\sim {\cal N}(0,\sigma^2_{sf})$. We assume $\kappa = 3.7$, $\sigma_{sf}^2 = 5$, and coherence block length $S= 500$.\footnotemark{}
\footnotetext{This coherence block can, for example, have the dimensions of $T_c =5\,{\rm{ms}}$ and $W_c = 100\,{\rm{kHz}}$.}

Statistical channel inversion power control is applied to both uplink pilot transmission and the ``virtual'' uplink payload power used in the M-MMSE precoder construction, i.e., $p_{lk} = \tau_{lk} =\frac{\rho}{d_l\left({\bf z}_{lk}\right)}$~\cite{bjornson2014massive}. Thus the average uplink SNR per antenna and user at its serving BS is constant: $\Eset\{p_{lk}\left\|{\bf h}_{llk}\right\|^2\}/(M\sigma^2) = \rho/{\sigma^2}$. This is a simple but effective policy to avoid near-far effects. For downlink payload data transmission, equal power allocation is used for each user, i.e., $\varrho_{lk}=P_{max}$. In the simulations, $\rho/{\sigma^2}$ is set to $0\,$dB to allow for decent uplink channel estimation accuracy, and $P_{max}$ is selected to make the cell edge SNR (without shadowing) equal to $-3\,$dB in the downlink.

\psfrag{Approximation beta=7 blablabla}{\Large {Approximation $\beta = 4$}}
\psfrag{data2}{\Large {Approximation $\beta = 7$}}
\psfrag{data3}{\Large {Approximation $\beta = 3$}}
\psfrag{data4}{\Large {Approximation $\beta = 1$}}
\psfrag{data5}{\Large {Simulation $\beta = 7$}}
\psfrag{data6}{\Large {Simulation $\beta = 4$}}
\psfrag{data7}{\Large {Simulation $\beta = 3$}}
\psfrag{data8}{\Large {Simulation $\beta = 1$}}
\psfrag{Number of Antennas}[][cb]{\Large {Number of Antennas}}
\psfrag{Achievable sum rate per cell (bit/s/Hz)}[][]{\Large{Achievable sum SE per cell (bit/s/Hz)}}
\psfrag{10}[][]{\Large {10}}
\psfrag{50}[][]{\Large {50}}
\psfrag{100}[][]{\Large {100}}
\psfrag{200}[][]{\Large {200}}
\psfrag{300}[][]{\Large {300}}
\psfrag{400}[][]{\Large {400}}
\psfrag{500}[][]{\Large {500}}
\psfrag{0}[][l]{\Large {0}}
\psfrag{10}[][l]{\Large {10}}
\psfrag{20}[][l]{\Large {20}}
\psfrag{30}[][l]{\Large {30}}
\psfrag{40}[][l]{\Large {40}}
\psfrag{50}[][l]{\Large {50}}
\psfrag{60}[][l]{\Large {60}}
\psfrag{70}[][l]{\Large {70}}
\psfrag{80}[][l]{\Large {80}}
\psfrag{90}[][l]{\Large {90}}
\psfrag{100}[][l]{\Large {100}}
\psfrag{120}[][l]{\Large {120}}
\psfrag{K=30}{\Large {$K=30$}}
\psfrag{K=10}{\Large {$K=10$}}
\begin{figure}[t]
\centering
\scalebox{0.44}{\includegraphics*{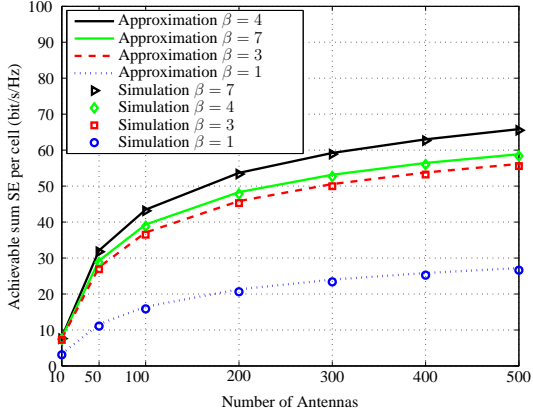}}
\caption{Achievable sum SE as a function of $M$, for $\beta\in\{1,3,4,7\}$ and $K=10$.}
\label{inst_determ}
 \vspace{-2ex}
\end{figure}

To verify the accuracy of the large-scale approximations, 10000 independent Monte-Carlo channel realizations for small scale fading are generated to numerically calculate the downlink achievable SE in~(\ref{rate_dl}). The numerical result and its approximation from Theorem~\ref{theorem4} are shown in Fig.~\ref{inst_determ}. As seen from the figure, the achievable sum SE increases monotonically with $\beta$ for the considered range of values. This is due to the following two properties. Firstly, a larger $\beta$ results in a lower level of pilot contamination, contributes to a higher channel estimation accuracy, and thereby increases the achievable SE. Secondly, a larger $\beta$ makes it possible for the BS to estimate more channel directions to users in other cells and design the precoders to reduce the interference caused to these other-cell users. However, it should be notice that a larger $\beta$ results in a larger pre-log loss, and thus the achievable SE may first increase and then decrease as $\beta$ grows when $S$ is small. Fig.~\ref{inst_determ} also shows that the numerical results and the large-scale approximations match very well, even for relatively small $M$ and small $K$.
\psfrag{K=30}{\Large {$K=30$}}
\psfrag{K=10}{\Large {$K=10$}}
\psfrag{Number of Antennas}[][cb]{\Large {Number of Antennas}}
\psfrag{Achievable sum rate per cell (bit/s/Hz)}[][]{\Large{Achievable sum SE per cell (bit/s/Hz)}}
\psfrag{500}[][]{\Large {500}}
\psfrag{200}[][]{\Large {200}}
\psfrag{300}[][]{\Large {300}}
\psfrag{400}[][]{\Large {400}}
\psfrag{100}[][]{\Large {100}}
\psfrag{120}[][]{\Large {120}}
\psfrag{140}[][]{\Large {140}}
\psfrag{160}[][]{\Large {160}}
\psfrag{180}[][]{\Large {180}}
\psfrag{M-MMSE: K=30 aaa}{\Large {M-MMSE: $K=30$}}
\psfrag{data5}{\Large {M-MMSE: $K=10$}}
\psfrag{data2}{\Large {M-ZF: $K=30$}}
\psfrag{data6}{\Large {M-ZF: $K=10$}}
\psfrag{data3}{\Large {S-MMSE: $K=30$}}
\psfrag{data7}{\Large {S-MMSE: $K=10$}}
\psfrag{data4}{\Large {MF: $K=30$}}
\psfrag{data8}{\Large {MF: $K=10$}}

\begin{figure}[t]
\centering
\scalebox{0.44}{\includegraphics*{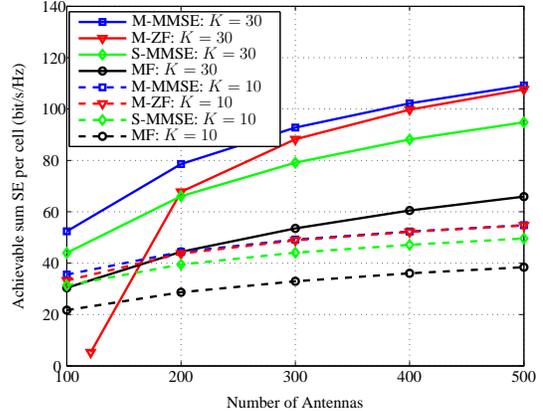}}
\caption{Achievable sum SE of M-MMSE, M-ZF, S-MMSE and M-MF for $\beta =4$.}
\label{sumrate_fullpower}
 \vspace{-3ex}
\end{figure}
To show the advantages of our M-MMSE precoder over other schemes, simulation results for the matched filter (MF) from~\cite{Marzetta10}, the multi-cell ZF (M-ZF) precoding from~\cite{bjornson2014massive}, and the S-MMSE precoding from~\cite{Hoydis2013} are provided for comparison in Fig.~\ref{sumrate_fullpower}. The same power allocation and normalization from~(\ref{m_precoder}) are applied for all precoders. Note that $M-\beta K>0$ is needed for the M-ZF scheme, thus the minimum value of $M$ for the M-ZF is $\beta K + 1$. We provide the results for $\beta=4$ in Fig.~\ref{sumrate_fullpower}. As seen from the figure, the MF scheme achieves the lowest performance since it does not actively suppress any interference. Compared with the S-MMSE, our M-MMSE achieves a notable SE gain and the gain increases as $K$ grows. For $M = 200$, the SE of M-MMSE is 10\% higher than that of S-MMSE with $K=10$, and the gain increases to 20\% with $K=30$. The advantage of the M-MMSE over the M-ZF is relatively small when $K=10$, but it becomes significant as $K$ grows. The performance gains over the S-MMSE and the M-ZF increase as $\beta$ grows as well, and the simulations results are provided in~\cite{xueru2015} due to the space limitation. Moreover, the M-ZF can sometimes achieve very low SE for small $M$, while our M-MMSE always seems to achieve good performance. In addition, the computational complexity of the M-MMSE is the same as for the M-ZF. Thus in general, our M-MMSE precoder is the better choice if high system SE is desirable.

\begin{figure} [t]
\centering
\scalebox{0.44}{\includegraphics*{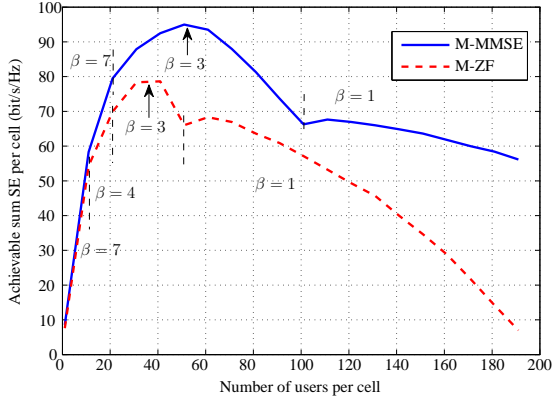}} 
\caption{Achievable sum SE as a function of $K$ with $M=200$.}
\label{sumrate_K}
 \vspace{-2ex}
\end{figure}

To show the ability of our M-MMSE precoder to handle different number of users, the SE is shown in Fig.~\ref{sumrate_K} as a function of $K$ for $M=200$. For each $K$, $\beta \in \{1,3,4,7\}$ are considered, and for each set of $K$ and $\beta$, we use the large-scale approximation given in Theorem~\ref{theorem4} to obtain the sum SE for the M-MMSE precoder. Fig.~\ref{sumrate_K} shows the SE with the $\beta \in \{1,3,4,7\}$ that that gives the highest SE for each $K$. The result for M-ZF is provided for comparison. As seen from the figure, the optimal value of $\beta$ changes for different $K$. Taking the M-MMSE precoder as an example, $\beta=7$ gives the highest SE for $K \leq 20$, then $\beta=3$ is desirable for a wide range of values of $K$, and finally the best choice is $\beta=1$ for $K \geq 100$ in order to reduce the pilot overhead. The maximal SE is achieved by $\beta=3$ for both precoders, but the corresponding $K$ and SE are higher for M-MMSE. It is notable that our M-MMSE can achieve high SE even for hundreds of users, while the performance of the M-ZF degrades drastically when $K > 70$, mainly since the array gain $M-\beta K$ is much smaller than $M$ in this range. Therefore, our M-MMSE can guarantee a much higher performance than the M-ZF in a crowded network.
\psfrag{Number of users per cell}[][]{\Large{Number of users per cell}}
\psfrag{Achievable sum SE per cell (bit/s/Hz)}[][]{\Large{Achievable sum SE per cell (bit/s/Hz)}}
\psfrag{beta=3}{\Large{$\beta=3$}}
\psfrag{beta=1}{\Large{$\beta=1$}}
\psfrag{beta=7}{\Large{$\beta=7$}}
\psfrag{beta=4}{\Large{$\beta=4$}}
\psfrag{M-MMSE}{\Large {M-MMSE}}
\psfrag{M-ZF}{\Large {M-ZF}}

\begin{figure}[t]
\centering
\scalebox{0.44}{\includegraphics*{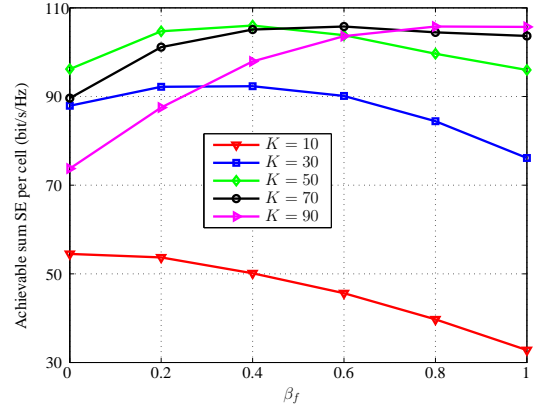}}
\caption{Achievable sum SE as a function of $\beta_{f}$ with $M=200$.}
\label{sumrate_betaF}
 \vspace{-3ex}
\end{figure}

Finally we consider the refined pilot reuse scheme proposed in~\cite{Atzeni2015icc} where pilot reuse is only applied at the cell edges. This reuse policy selects $B = K (\beta_f + (1-\beta_f)\beta)$ pilot sequences, and assign the same $\beta_f K $ pilot sequences to the $\beta_f K$ users that are closest to their serving BSs. The other $K(1-\beta_f)\beta$ pilots are assigned to the remaining users according to the same symmetric reuse policy as before. The refined scheme reduces to the symmetric reuse policy for $\beta_f=0$, while $\beta_f=1$ means that the same pilots are used in all cells. The idea of this scheme is to mitigate pilot contamination while limiting the pilot overhead. We apply this scheme to our M-MMSE precoder, and show the performance in Fig.~\ref{sumrate_betaF} for $M=200$ and different $K$. For each $K$ and $\beta_f$, we choose the highest SE from what we obtained from $\beta \in \{1,3,4,7\}$. As seen from the figure, $\beta_f =0$ is preferred when $K$ is small, which means that pilot reuse should also be applied for everyone. However, $\beta_f >0$ is beneficial for larger $K$ values since there will likely be cell center users that are not sensitive to pilot contamination, which can be exploited as in~\cite{Atzeni2015icc} to reduce the pilot overhead. However, if we look at the highest SE for each $\beta_f$ among the considered $K$ and $\beta$, the increase of $\beta_f$ is not as obvious as it is for M-ZF shown in~\cite{Atzeni2015icc}, thanks to the strong ability of our M-MMSE to actively suppress inter-cell interference.
\psfrag{0.2}[][]{\Large {0.2}}
\psfrag{0.4}[][]{\Large {0.4}}
\psfrag{0.6}[][]{\Large {0.6}}
\psfrag{0.8}[][]{\Large {0.8}}
\psfrag{1}[][]{\Large {1}}

\psfrag{110}[][]{\Large {110}}
\psfrag{130}[][]{\Large {130}}

\psfrag{K=10 bla}{\Large {$K=10$}}
\psfrag{K=30}{\Large {$K=30$}}
\psfrag{K=50}{\Large {$K=50$}}
\psfrag{K=70}{\Large {$K=70$}}
\psfrag{K=90}{\Large {$K=90$}}
\psfrag{betaF}{\Large{$\beta_f$}}
\psfrag{sum SE}[][l]{\Large {Achievable sum SE per cell (bit/s/Hz)}}

\section{Conclusions} \label{conclusions}
In this paper, a multi-cell MMSE precoder is proposed based on uplink-downlink duality, and an approximative SINR expression is derived which is tight in the large-system limit. Compared with conventional single-cell precoders, that only make use of the estimated channel directions within the serving cell, our multi-cell MMSE scheme utilizes all channel directions that can be estimated at the BS to also mitigate inter-cell interference. Numerical results show that the proposed precoder brings a very promising sum SE gain over the single-cell MMSE and the gain becomes significant as $K$ or $\beta$ grows. Compared to the M-ZF, our M-MMSE precoder can maintain high and stable SE for very large $K$. Furthermore, the SINR approximation is easy to compute and shown to be accurate even for small system dimensions.

\vspace{-1ex}
\bibliographystyle{IEEEtran}
\linespread{1.0}\selectfont
\bibliography{mmse_gc_dl}

\begin{thebibliography}{10}
\providecommand{\url}[1]{#1}
\csname url@samestyle\endcsname
\providecommand{\newblock}{\relax}
\providecommand{\bibinfo}[2]{#2}
\providecommand{\BIBentrySTDinterwordspacing}{\spaceskip=0pt\relax}
\providecommand{\BIBentryALTinterwordstretchfactor}{4}
\providecommand{\BIBentryALTinterwordspacing}{\spaceskip=\fontdimen2\font plus
\BIBentryALTinterwordstretchfactor\fontdimen3\font minus
  \fontdimen4\font\relax}
\providecommand{\BIBforeignlanguage}[2]{{%
\expandafter\ifx\csname l@#1\endcsname\relax
\typeout{** WARNING: IEEEtran.bst: No hyphenation pattern has been}%
\typeout{** loaded for the language `#1'. Using the pattern for}%
\typeout{** the default language instead.}%
\else
\language=\csname l@#1\endcsname
\fi
#2}}
\providecommand{\BIBdecl}{\relax}
\BIBdecl

\bibitem{Marzetta10}
T.~L. Marzetta, ``Noncooperative cellular wireless with unlimited numbers of
  base station antennas,'' \emph{IEEE Trans. Wireless Communications}, vol.~9,
  no.~1, pp. 3590--3600, Nov. 2010.

\bibitem{Rusek13}
F.~Rusek, D.~Persson, K.~L. Buon, E.~G. Larsson, T.~L. Marzetta, O.~Edfors, and
  F.~Tufvesson, ``Scaling up {MIMO}: Opportunities and challenges with very
  large arrays,'' \emph{IEEE Trans. Signal Process.}, vol.~30, no.~1, pp.
  40--60, Jan. 2013.

\bibitem{Larsson14}
E.~G. Larsson, O.~Edfors, F.~Tufvesson, and T.~L. Marzetta, ``Massive {MIMO}
  for next generation wireless systems,'' \emph{IEEE Commun. Mag.}, vol.~52,
  no.~2, pp. 186--195, Feb. 2014.

\bibitem{bjornson2014massive}
E.~Bj{\"o}rnson, E.~G. Larsson, and M.~Debbah, ``Massive {MIMO} for maximal
  spectral efficiency: how many users and pilots should be allocated?''
  \emph{IEEE Trans. Wireless Commun}, Dec. 2014, submitted.

\bibitem{bjornson2013massive}
E.~Bj{\"o}rnson, J.~Hoydis, M.~Kountouris, and M.~Debbah, ``Massive {MIMO}
  systems with non-ideal hardware: energy efficiency, estimation, and capacity
  limits,'' \emph{IEEE Trans. Inf. Theory}, vol.~60, no.~11, pp. 7112--7139,
  Nov. 2014.

\bibitem{Hien13}
H.~Q. Ngo, E.~G. Larsson, and T.~L. Marzetta, ``Energy and spectral efficiency
  of very large multiuser {MIMO} systems,'' \emph{IEEE Trans. Commun.},
  vol.~61, no.~4, pp. 1436--1449, Apr. 2013.

\bibitem{Marzetta2006}
T.~L. Marzetta, ``How much training is required for multiuser {MIMO}?'' in
  \emph{Proc. IEEE Asilomar Conference on Signals, Systems and Computers}, Oct.
  2006, pp. 359--363.

\bibitem{Hoydis2013}
J.~Hoydis, S.~ten Brink, and M.~Debbah, ``Massive {MIMO} in the {UL/DL} of
  cellular networks: how many antennas do we need?'' \emph{IEEE J. Sel. Areas
  Commun.}, vol.~31, no.~2, pp. 160--171, Feb. 2013.

\bibitem{Yang2013performance}
H.~Yang and T.~L. Marzetta, ``Performance of conjugate and zero-forcing
  beamforming in large-scale antenna systems,'' \emph{IEEE J. Sel. Areas
  Commun.}, vol.~31, no.~2, pp. 172--179, Feb. 2013.

\bibitem{Jose2011}
J.~Jose, A.~Ashikhmin, T.~L. Marzetta, and S.~Vishwanath, ``Pilot contamination
  and precoding in multi-cell {TDD} systems,'' \emph{IEEE Trans. Wireless
  Commun.}, vol.~10, no.~8, pp. 2640--2651, Aug. 2011.

\bibitem{xueru2015UL}
X.~Li, E.~Bj{\"o}rnson, E.~G. Larsson, S.~Zhou, and J.~Wang, ``A multi-cell
  {MMSE} detector for massive {MIMO} systems and new large system analysis,''
  in \emph{Proc. IEEE GLOBECOM}, 2015, submitted.

\bibitem{Wagner2012}
S.~Wagner, R.~Couillet, M.~Debbah, and D.~Slock, ``Large system analysis of
  linear precoding in correlated {MISO} broadcast channels under limited
  feedback,'' \emph{IEEE Trans. Inf. Theory}, vol.~58, no.~7, pp. 4509--4537,
  Jul. 2012.

\bibitem{xueru2015}
X.~Li, E.~Bj{\"o}rnson, E.~G. Larsson, and J.~Wang, ``Massive {MIMO} with
  multi-cell {MMSE} processing: exploiting all pilots for interference
  suppression,'' \emph{IEEE Trans. Wireless Commun.}, Apr. 2014, submitted.

\bibitem{Atzeni2015icc}
I.~Atzeni, J.~Arnau, and M.~Debbah, ``Fractional pilot reuse in massive {MIMO}
  systems,'' in \emph{Proc. IEEE ICC, Workshops 23}, Jun. 2015.

\end{thebibliography}

\end{document}